\documentclass[preprint,12pt]{elsarticle}

\usepackage{hyperref}
\usepackage{cleveref}
\usepackage{amssymb}

\usepackage{geometry}
\usepackage{lineno}
\usepackage{bm}
\usepackage{siunitx}

\begin{document}

\begin{frontmatter}



\title{Dynamics of a toroidal bubble on a cylinder surface with an application to geophysical exploration}

\author[1]{Shuai Li\corref{cor1}}
\author[1,2]{Andrea Prosperetti}
\author[1]{Devaraj van der Meer}

\cortext[cor1]{Corresponding author: s.li@utwente.nl; lishuai@hrbeu.edu.cn}

\address[1]{Physics of Fluids Group, Max-Planck Center for Complex Fluid Dynamics, MESA+ Research Institute, J.M. Burgers Center for Fluid Dynamics, Department of Science \& Technology, University of Twente, P.O. Box 217, 7500 AE Enschede, The Netherlands}
\address[2]{Department of Mechanical Engineering, University of Houston, TX 77204-4006, USA}

\begin{abstract}
During the operation of a seismic airgun source, a certain amount of compressed high-pressure air is released from the airgun chamber into the surrounding water, generating an expanding toroidal bubble attached to the airgun-body. The subsequent oscillations of the bubble generate low-frequency pressure waves, which are used to map the ocean subbottom, e.g., to locate oil and gas reserves. The bubble dynamic behavior and the emitted pressure waves are inevitably influenced by the airgun-body. However, the bubble-airgun-body interaction is far from well understood. This paper investigates the strong interaction between a long cylinder and an attached toroidal bubble via hundreds of boundary integral simulations, aiming to provide new physical insights for airgun-bubble dynamics. Firstly, the overall physical phenomena are discussed and three types of bubble collapse patterns are identified, namely (i) upward jetting due to gravity, (ii) annular jet toward the cylinder body and (iii) weak/no jet. Thereafter, we investigate the effects of the cylinder radius, initial bubble pressure and Froude number on the bubble oscillation period and the pressure wave induced by the bubble. At last, the impact of a cylinder on a Sercel type airgun-bubble is discussed with a particular focus on the spectrum of the pressure waves.

\end{abstract}

\begin{keyword}
Bubble dynamics  \sep Seismic airgun source \sep Boundary integral method  \sep Geophysical exploration \sep Fluid-structure interaction
\end{keyword}

\end{frontmatter}


\section{Introduction}
\label{S:1}
Bubble dynamics is a long-standing research topic owing to its rich behavior and many applications \cite{Lohse-prf,Dollet2019,Prosperetti2017,Badve2015,Pawar2017}. In the fields of hydraulic machinery and ship propulsion, cavitation erosion is a serious problem to submerged structures and cavitation bubbles are also the main cause of broadband noise and vibration \cite{JiB2015}. In geophysical exploration \cite{Watson2019,Chelminski}, seismic surveys utilize airguns to produce powerful pressure waves with a broad low-frequency spectrum. The specific interest in low-frequency waves is motivated by their ability to penetrate deep into the seabed.

This study aims to explore the interaction between the bubble and the airgun-body. In most of the existing works \cite{Graaf2014,Ziolkowski1970,Ziolkowski1982,ZhangS2017,Khodabandeloo2018}, the Rayleigh-Plesset equation \cite{Rayleigh,Plesset1949} has been used for airgun modeling. However, the spherical symmetry assumption which underlies the Rayleigh-Plesset equation is inappropriate for airgun bubbles because the action of gravity causes highly non-spherical deformation of such large-scale bubbles \cite{Li-IJMF}. For example, Benjamin \& Ellis \cite{Benjamin1966} observed that an upward liquid jet forms during the oscillation of a bubble in a gravity field. It is also well known that a collapsing bubble moves towards a rigid wall due to the image effect \cite{Zeng-Ohl,Blake1987,Brujan2018}, and in the process produces a liquid jet that impacts the wall. More generally, the non-spherical bubble dynamics is highly dependent on boundary conditions. The interactions between a single bubble and many kinds of boundaries have been investigated, such as free surfaces \cite{Koukouvinis-free,Kannan2018,Kannan2020}, elastic membranes \cite{Klaseboer2006}, suspended objects \cite{Li-jcp,Borkent}, floating structures \cite{Klaseboer2005,Cui-jfm}, other bubbles \cite{Bremond2006,Bremond-prl}, etc. Nevertheless, the presence of the airgun-body itself imposes an additional important geometric boundary condition which is quite different from those mentioned above. 

The majority of seismic airguns have a cylindrical shape and contain a chamber filled with highly compressed air. The sleeve-type airgun has an annular port in the middle of the airgun-body \cite{mayzes1994monoport}. Upon firing, a growing toroidal bubble is generated as the compressed air is released from the airgun chamber into the surrounding water. The toroidal bubble is inevitably influenced by the airgun-body. However, we have little knowledge of the interaction between the airgun-body and an attached toroidal bubble. In some spherical bubble models, the airgun-body effect is approximated by assuming the airgun is a solid core inside the bubble and only influences the gas volume which is now calculated by subtracting the airgun volume from the bubble volume $4\pi R^3/3$, where $R$ is the radius of the bubble \cite{Graaf2014,Schulze-Gattermann}. This method may be suitable for very short airguns. However, many airguns are relatively long and cannot be fully enclosed by the bubble. Therefore, a better understanding of the airgun-bubble requires a fully coupled bubble-structure modeling. 

It has been demonstrated that the viscosity and compressibility of water do not play a significant role in the first and second oscillating cycles of a typical airgun-bubble \cite{Graaf2014,Ziolkowski1970,Langhammer-visosity}. Hence, we establish a numerical model for bubble-cylinder interaction based on potential flow theory implemented with a boundary integral (BI) method. The cylinder is positioned vertically and the bubble constitutes a half-toroid attached to the airgun-body. With this assumption, the problem can be solved in an axisymmetric configuration. As mentioned above, this model is very suitable for simulating bubbles generated by sleeve-type airguns. For other types of airguns, several ports are evenly arranged in a circle around the longitudinal axis of the airgun. The generated sub-bubbles coalesce into a bigger toroidal bubble within a short time \cite{Langhammer1996,Langhammer-phd}, which makes the present model suitable for this type of airgun as well.

This paper is organized as follows. The physical problem and numerical setup are introduced in Section \ref{S:2}. The numerical validation is given in Section \ref{S:4-1} and more discussion is presented in Sections \ref{S:4-2}-\ref{S:4-4}. Here, we focus on the bubble collapse pattern, oscillation period and pressure wave as functions of the governing parameters (the cylinder radius, initial bubble pressure, and Froude number). In Section \ref{S:5}, we use the parameters of a real airgun to investigate the impact of the cylinder on the emitted pressure wave and spectrum. Conclusions are drawn in Section \ref{S:con}.

\section{Problem statement and numerical setup}
\label{S:2}

\begin{figure}[htbp]
	\centering\includegraphics[width=8cm]{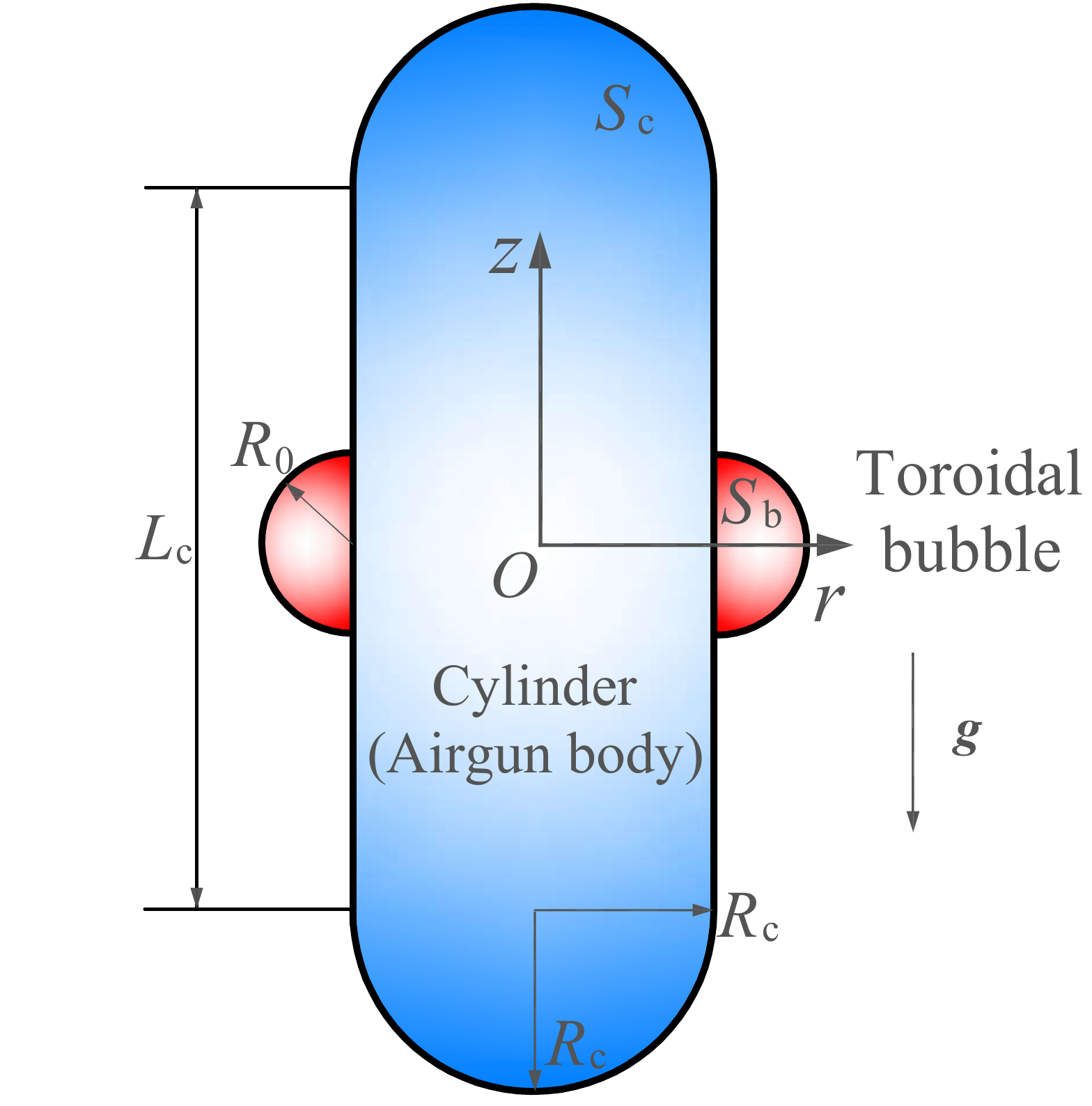}
	\caption{Sketch of the physical problem for the interaction between a cylinder and an attached toroidal bubble.}\label{Fig:sketch}
\end{figure}

Figure \ref{Fig:sketch} shows a sketch of the physical problem. The airgun-body is simplified as a cylinder with two hemispherical ends and the cylinder radius is denoted by $R{\rm_c}$. The bubble at the initial moment is a half-toroid with its radius and pressure denoted by $R_0$ and $P_0$ (see Figure \ref{Fig:sketch}). The longitudinal axis of the cylinder is parallel to the gravity direction, leading to an axisymmetric configuration of the system. A cylindrical coordinate system $(r, z)$ is defined with the origin $O$ placed at the center of the cylinder and the $z$-axis pointing upward. 

The Reynolds number associated with a typical airgun-bubble is over $10^7$ and the Mach number is much smaller than one \cite{Cox}. Furthermore, the scale of the bubble is very large (the maximum radius of the bubble is in order of $O$(1 m)), thus the details of the contact line motion over the cylinder body have no dynamic significance. Therefore, the viscosity and compressibility of the liquid can be safely neglected in this problem. With these assumptions, the velocity field can be assumed to be irrotational, which permits us to express it as the gradient of the potential
\begin{equation}
\textbf{\textit{u}}^* = \nabla \varphi^*,
\end{equation}
where by the condition of incompressibility, the velocity potential satisfies the Laplace's equation
\begin{equation}
\nabla^2{\varphi^*}=0,
\end{equation}
where the asterisk * denotes a dimensionless quantity. 

Instead of solving the Laplace's equation directly, we can use Green's second identity to transform it into the following boundary integral equation:
\begin{equation} 
\theta(\textbf{\textit{r}}^*){\varphi^*}(\textbf{\textit{r}}^*) = \int\left[ {\frac{\partial{\varphi^*}(\textbf{\textit{q}}^*)}{\partial{n}}\frac{1}{|\textbf{\textit{r}}^*-\textbf{\textit{q}}^*|}-{\varphi^*}(\textbf{\textit{q}}^*)\frac{\partial{}}{\partial{n}}}\left( \frac{1}{|\textbf{\textit{r}}^*-\textbf{\textit{q}}^*|}\right) \right] {\rm d}S{\rm_{q^*}},\label{Equation:BI}
\end{equation}
where $S_{{\rm_{q^*}}}$ includes the bubble surface $S_{\rm b}$ and cylinder surface $S_{\rm c}$;   $\textbf{\textit{r}}^*$ is the control point, $\textbf{\textit{q}}^*$ is the integral point, $\theta$ is the solid angle, and $\partial/\partial n$ is the normal derivative at the boundary. A standard boundary integral (BI) method  is used to solve the above equation \cite{Wang1996,Bouwhuis2016,Oguz1993,Zhang2015b}.

For the nondimensionalization, the maximum equivalent bubble radius $R_{\rm m}$, liquid density $\rho$ and ambient hydrostatic pressure at the initial bubble center $P_\infty$ are chosen as basic quantities. This implies the following length, velocity, time and pressure scales: $R_{\rm m}$, $\sqrt{\frac{P_\infty}{\rho}}$, $R_{\rm m}\sqrt{\frac{\rho}{P_\infty}}$, and $P_\infty$. This choice sets the dimensionless maximum volume of the bubble to $V_{\rm m}^*=4\pi/3$. Note that there is static pressure gradient in the liquid although we don't consider explicitly the distance of the bubble from the free surface. 

Since the cylinder is stationary in our model, the boundary condition on $S_{\rm c}$ is given by
\begin{equation}
\nabla \varphi^*\cdot\textbf{\textit{n}}=0,\label{Equation:wall-BC}
\end{equation}
where $\textbf{\textit{n}}$ is the unit normal vector pointing out of the fluid domain.

In every time step, the position of the bubble surface is updated via the kinematic boundary condition
\begin{equation} 
\frac{{\rm d}\textbf{\textit{r}}^*}{{\rm d}t^*}=\nabla{\varphi}^*,\label{Equation:KBC}
\end{equation}
and the velocity potential on $S_{\rm b}$ is updated by time-integrating the dynamic boundary condition on the bubble surface
\begin{equation} 
\frac{{\rm d}{\varphi^\ast}}{{\rm d}t^\ast}=\frac{1}{2}{|{\nabla \varphi^\ast}|^2}-P_{0}^\ast\left(\frac{V_0^\ast}{V^\ast} \right) ^\gamma-\frac{1}{Fr^2}z^\ast+1.\label{Equation:DBC}
\end{equation}
The Froude number $Fr$ is defined as
\begin{equation} 
Fr=\sqrt{\frac{P_\infty}{\rho gR_{\rm m}}},\label{Equation:Fr}
\end{equation}
and the adiabatic equation, $P_{\rm b}^*=P_{0}^\ast\left(\frac{V_0^\ast}{V^\ast} \right) ^\gamma$, is used to describe the pressure inside the bubble. $\gamma$ is the adiabatic index, which equals 1.4 for air. The surface tension is also neglected in Equation (\ref{Equation:DBC}) because the associated Weber number is over $10^6$.

For a specified initial bubble pressure $P_0^*$, the initial bubble volume $V_0^*$ can be derived from the energy conservation law \cite{Klaseboer2006}:
\begin{equation} 
V_0^*+\frac{P_0^*V_0^*}{\gamma-1}=V_{\rm m}^*+\frac{P_0^*V_{\rm m}^*}{\gamma-1}\left(\frac{V_0^*}{V_{\rm m}^*}   \right)^\gamma. \label{eq:V0}
\end{equation}

The initial radius of the bubble can be obtained by equating the volume of the half-toroid to the initial bubble volume
\begin{equation} 
\frac{4}{3}\pi R_0^{*3}+\pi^2R{\rm_c^*}R_0^{*2}=V_0^*. \label{eq:R0}
\end{equation}

Keep in mind that (since $V_{\rm m}^*=4\pi/3$) the initial volume (energy) of the bubble remains the same in all simulations for a specified $P_0^*$. Although the bubble shape becomes nonspherical due to the presence of the cylinder, we found that the maximum bubble volume $V_{\rm m}^*$ obtained in the simulation is insensitive to  $R{\rm_c^*}$.

There are two moving contact lines in the system, namely the two intersection lines between the bubble surface and the cylinder surface (see Figure \ref{Fig:sketch}). Both the boundary conditions on the bubble surface and the solid wall (free slip boundary condition) need to be imposed on these contact lines simultaneously. Following Ni et al \cite{NiBY} and Li et al \cite{Li-jcp}, we adopt the artifice of considering each contact line as a pair of coincident lines, one of them considered as belonging to the cylinder and the other one to the bubble. On the former we impose the vanishing of the velocity normal to the cylinder, 
Equation (\ref{Equation:wall-BC}). On the latter we impose the condition (Equation \ref{Equation:DBC}) that determines the potential. 

The far-field pressure of an acoustic monopole source is given by \cite{landau1987fluid}
\begin{equation}
P= \frac{\rho}{4\pi r}\frac{{\rm d^2}{V}}{{\rm d}t^{2}},\label{eq:pressure}
\end{equation}
where $r$ is the distance between the field point and the source. It is customary in the airgun community to set $r=$ 1 m to scale the acoustic pressure at any specific distance in inverse proportion to the actual distance from the source. In order to be able to apply this relation, we calculate $P^*\big|_{r^*=1} = \frac{1}{4\pi}\frac{{\rm d^2}{V^*}}{{\rm d}t^{*2}}$ from our computed bubble dynamics, unless stated otherwise.

In the above system, the bubble dynamics is governed by the cylinder size (the radius $R_{\rm c}^*$ and the length $L_{\rm c}^*$), the initial bubble pressure $P_0^*$ and gravity $Fr$. However, we find that the cylinder length $L_{\rm c}^*$ has a negligible influence on the bubble motion as long as it is larger than the maximum bubble diameter. Thus we only have three governing parameters. The influence of a short cylinder on a relatively larger bubble ($L_c<2R_{\rm m}$) is not discussed in this paper. We set $L_{\rm c}^*$ = 10 in the following numerical simulations, i.e., much larger than the maximum bubble radius $R_{\rm m}^*=1$ and equivalent to infinitely long for all practical purposes. 

Finally, we give the numerical procedure for the simulation of the bubble-cylinder interaction before the jet impact: (i) Initialization and nondimensionalization; (ii) Begin time stepping and set an adaptive time step \cite{Wang1996}; (iii) Solve Equation (\ref{Equation:BI}) and obtain the velocity on the bubble surface; (iv) Update the location of the bubble surface using Equation (\ref{Equation:KBC}); (v) Update the potential on the bubble surface using Equation (\ref{Equation:DBC}); (vi) Calculate the far-field pressure; (vii) Remesh the bubble surface and the cylinder surface using spline interpolation \cite{Zhang2015b,Pearson2004-jet}; (viii) Go back to (ii) until the jet impact.

\section{Results and discussions}
\label{S:4}

\subsection{Numerical validation}
\label{S:4-1}

\begin{figure}[htbp]
	\centering\includegraphics[width=16cm]{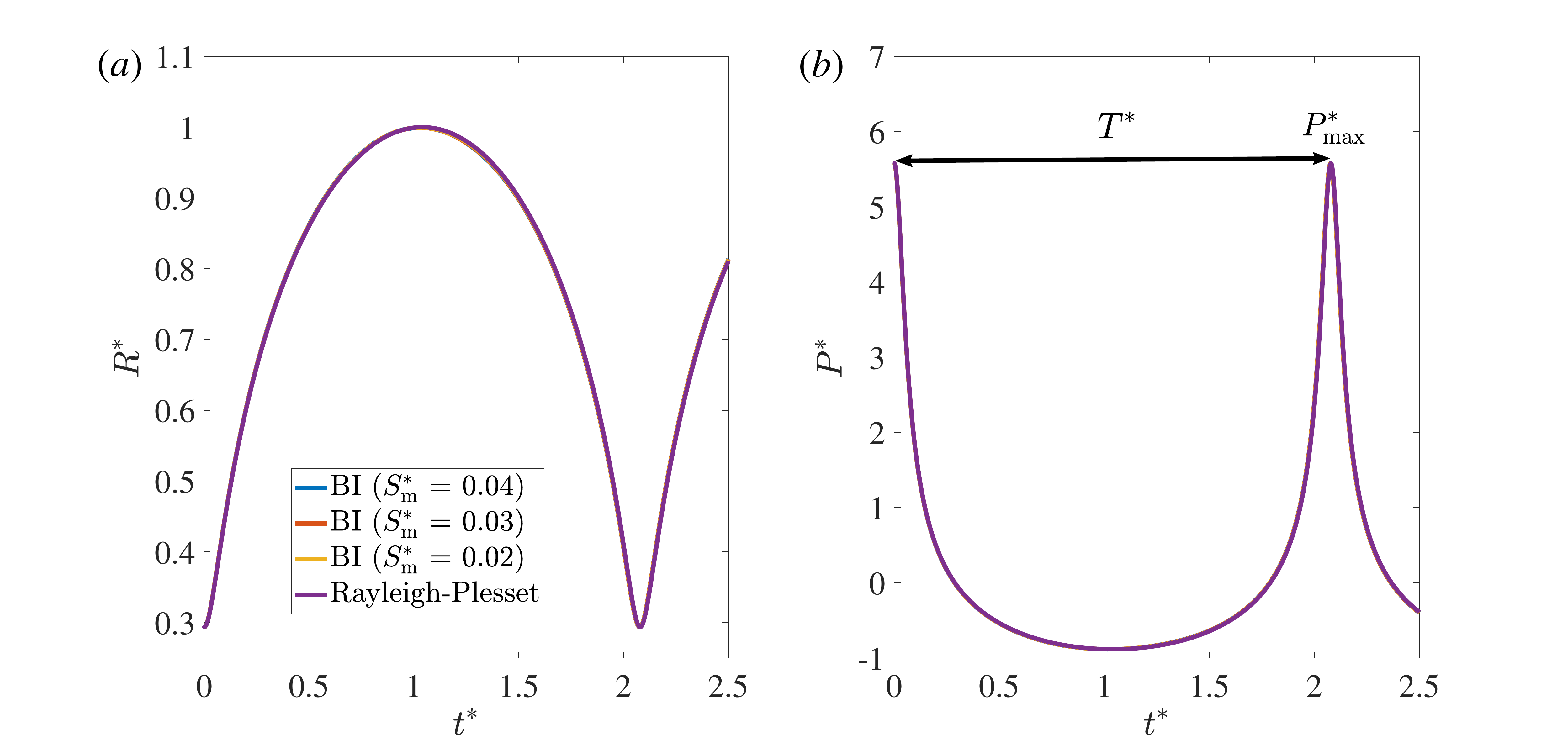}
	\caption{Comparison of (a) the equivalent bubble radius and (b) the pressure wave between numerical simulations and the analytical solution from Rayleigh-Plesset equation \cite{Rayleigh}. In the BI simulations, the bubble is interacting with a very thin cylinder ($R_{\rm c}^* = 0.002$) and different mesh sizes are used, i.e., $S_{\rm m}^*\rm = 0.04, 0.03, and\ 0.02$. }\label{Fig:verification1}
\end{figure}

To validate the code, we use three test problems. In the first one, we compare the numerical result with an analytical solution of the Rayleigh-Plesset equation \cite{Rayleigh,Plesset1949} for a case in which the cylinder is set to be very thin, $R_{\rm c}^* = 0.002$. We omit the effect of gravity in order to maximize the accuracy of the spherical bubble model. The initial dimensionless bubble radius and pressure are set to $R_0^* = 0.292$ and $P_0^* = 20$. The effect of the mesh size $S_{\rm m}^*$, defined as the dimensionless distance between two neighboring nodes on the bubble surface, is also tested. The element number varies during the simulation to keep $S_{\rm m}^*$ almost the same. Comparisons of the bubble radius and pressure wave between numerical simulations and the analytical solution are shown in Figure \ref{Fig:verification1}. As can be seen, the BI simulation results agree very well with the analytical result and the numerical result is independent of the mesh size. Two important quantities are defined in panel (b), namely, the bubble oscillation period $T^*$ and the maximum pressure induced by the bubble collapse $P_{\rm max}^*$, which are crucial to the spectrum of the pressure wave. 

\begin{figure}[htbp]
	\centering\includegraphics[width=7.5cm]{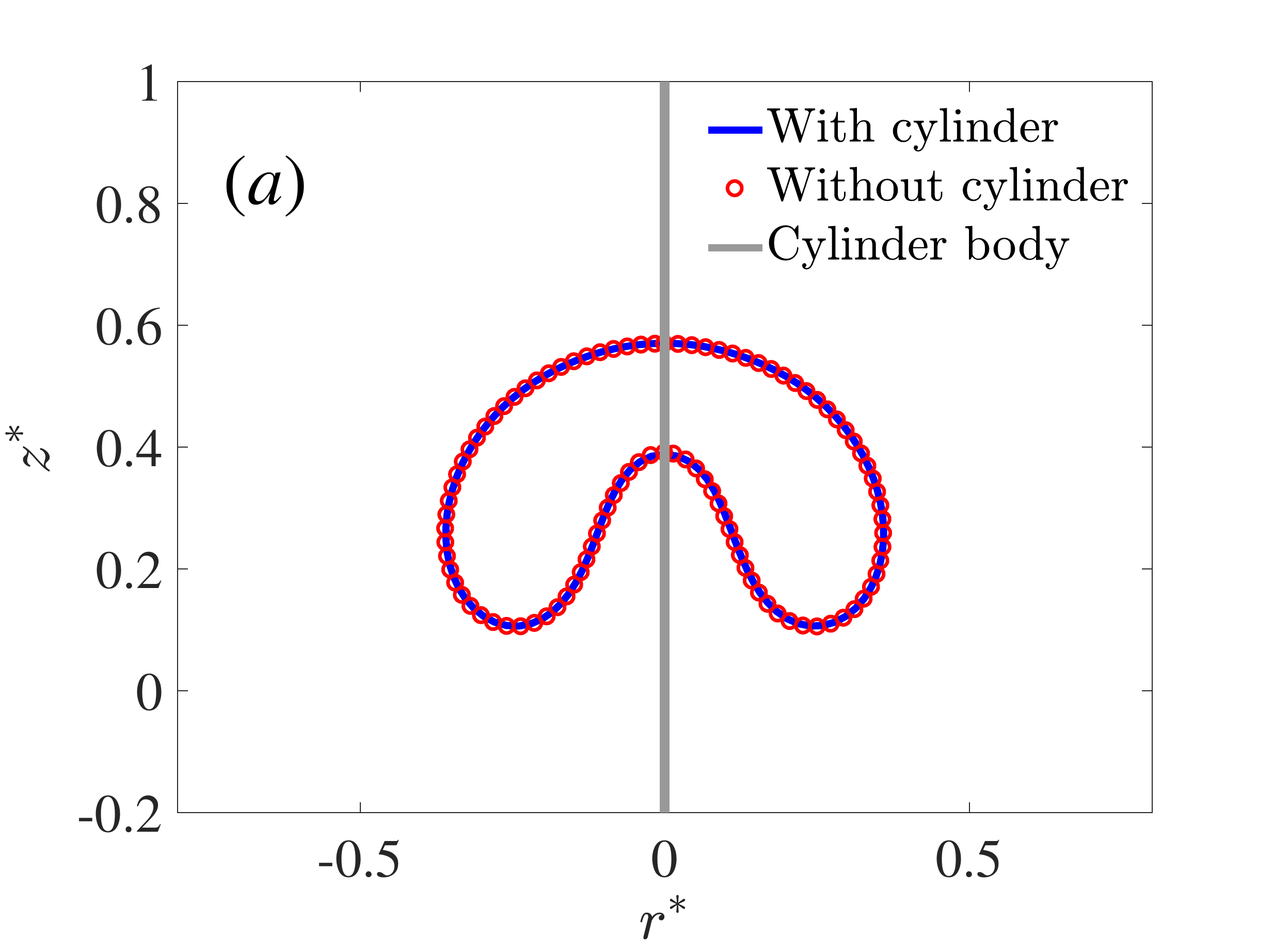}
	\centering\includegraphics[width=7.5cm]{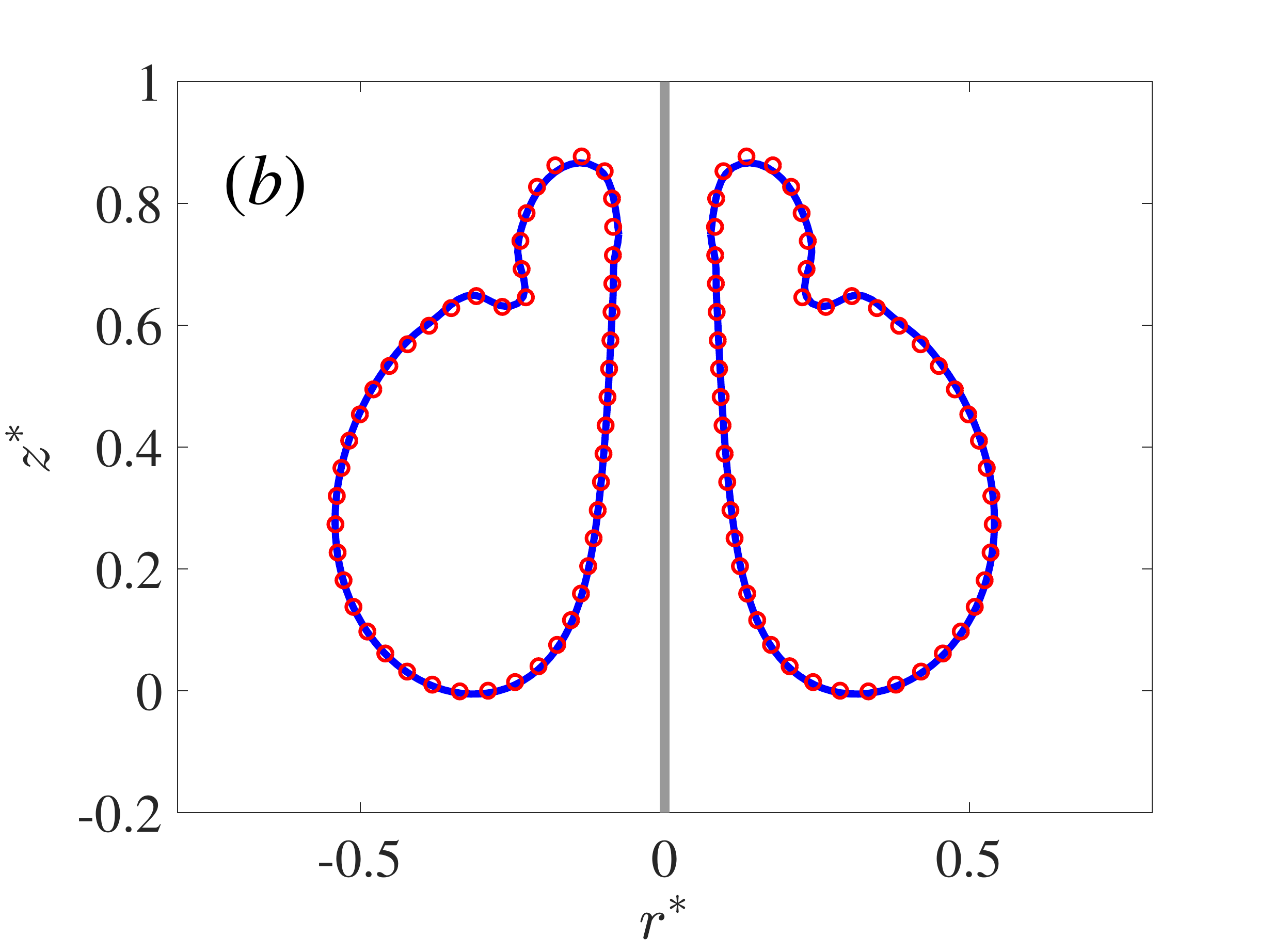}
	\caption{Comparison of bubble profiles under gravity ($Fr = 5$). Other parameters are kept the same as in Figure \ref{Fig:verification1}. The solid blue lines denote the simulation results of the bubble-cylinder interaction and the red circles denote the bubble motion without a cylinder. The dimensionless times are 2.08  (a) and 2.20 (b), respectively.}\label{Fig:verification2}
\end{figure}

As a further test of the contact line model, we include the effect of gravity ($Fr = 5$) in the previous simulation and compare the results with a simulation in which the cylinder is removed. In this case, an upward high-speed liquid jet forms during the collapse.  The computed nonspherical bubble profiles exhibit excellent agreement between the two simulations, as shown in Figure \ref{Fig:verification2}.

\begin{figure}[htbp]
	\centering\includegraphics[width=16cm]{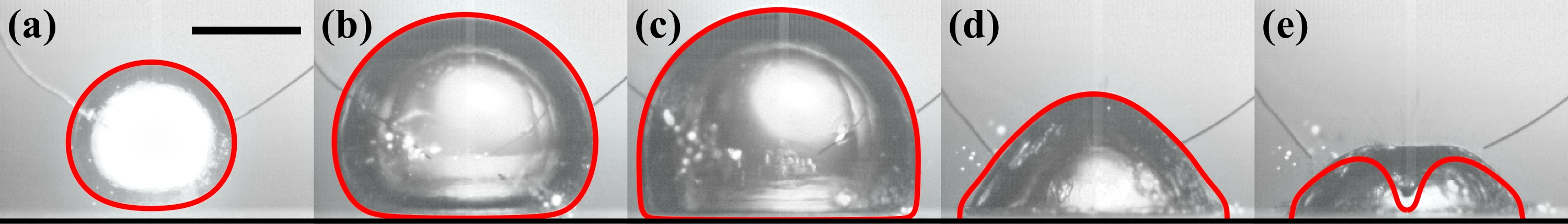}
	\caption{Comparison of the non-spherical bubble dynamics in contact with a rigid wall (side view) between the experimental observation and a boundary integral simulation (denoted by the red lines). In the experiment, the bubble is generated by the underwater electric discharge method (800V DC). The initial distance from the bubble center to the wall is 6.8 mm and the maximum equivalent bubble radius is about 12.3 mm. The times of each frame are 0.32, 0.89, 1.52, 2.47 and 2.59 ms, respectively. The black bar shows a 10 mm scale. The mesh size is set as $S_{\rm m}^* = 0.02$ in the simulation.}\label{Fig:exp-verification}
\end{figure}

As a final validation,  we compare the numerical simulation with the  results of an experiment conducted for this purpose, as shown in Figure \ref*{Fig:exp-verification}. Details of the method and of the experimental setup can be found in Cui et al \cite{Cui-jfm}. Briefly,  a bubble was generated by an underwater electric discharge (800 V dc) above a flat horizontal rigid wall. The initial distance of the bubble center from the wall was 6.8 mm and the maximum equivalent bubble radius $R_{\rm m}$ about 12.3 mm. The lower surface of the bubble approaches the wall during the expansion and becomes gradually flattened  (frames a-b). Frame (c) shows the moment of the maximum volume. During the bubble collapse stage (frames d-e), a high-speed liquid jet forms from the bubble top and impacts the wall. In the BI simulation, the liquid film between the bubble and the wall tends to rupture since the lubrication force is not included. To maintain the stability of the simulation, we need to remove the thin film when its thickness becomes smaller than the local meshsize (around frame c). Thereafter, a liquid-solid-gas contact line forms. Our numerical treatment of this contact line is accurate, as suggested by the fact that the evolution of the bubble surface is well reproduced by our model (frames d-e).

\begin{figure}[htbp]
	\centering\includegraphics[width=13.5cm]{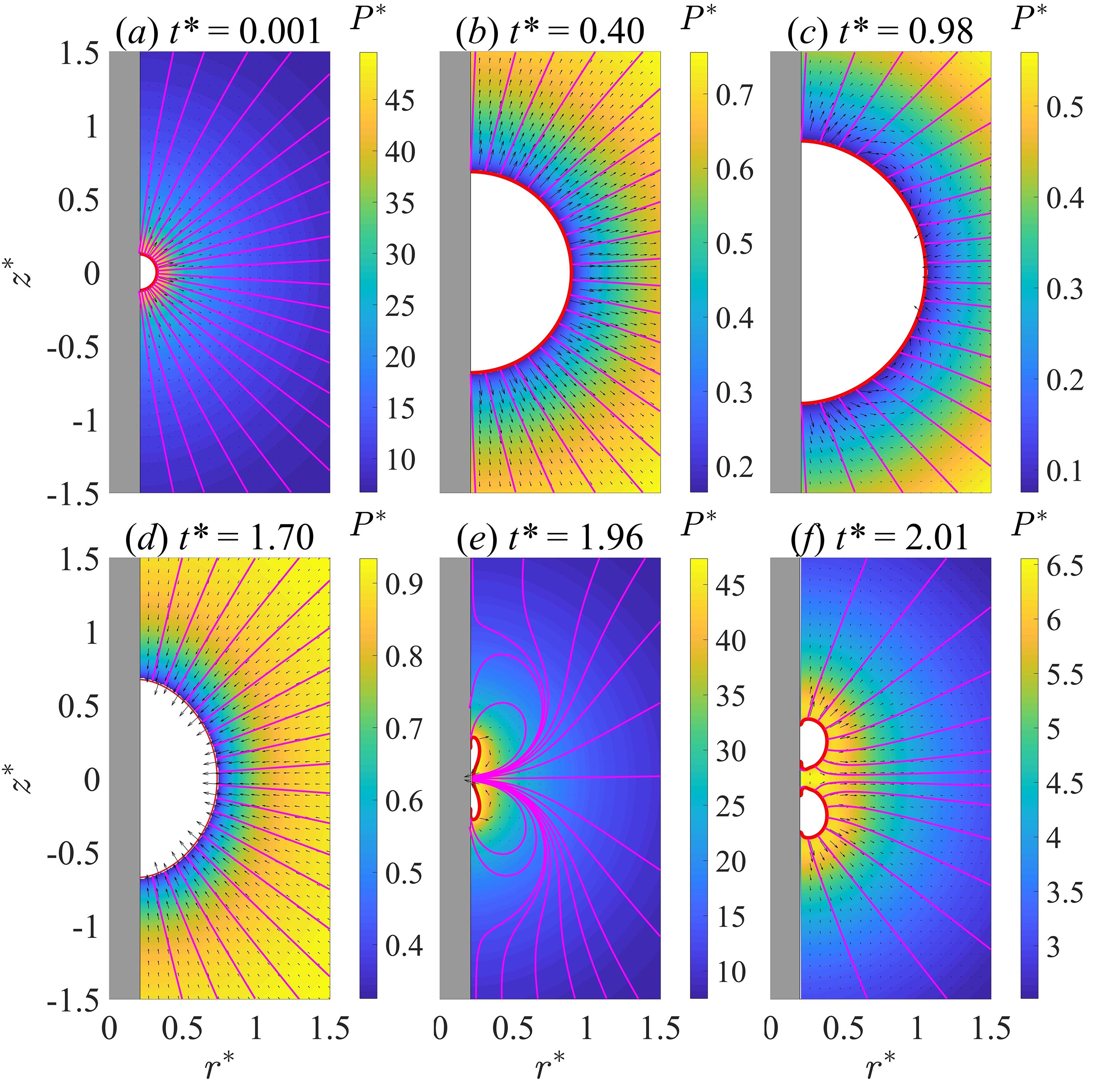}
	\caption{The dynamic behavior of an attached toroidal bubble on a cylinder surface for $Fr = \infty$, $R_{\rm c}^* = 0.2$, $P_0^*=50$ and $R_{0}^* = 0.1253$. In this and subsequent figures, the contours denote the pressure field, the arrows denote the velocity field and the solid magenta lines denote the streamlines.  The grey rectangle is the semi-section of the cylinder.}\label{Fig:4-1}
\end{figure}

\subsection{Overall physical phenomena}
\label{S:4-2}

We carried out hundreds of simulations for different parameter sets ($0 \leq R_c^* \leq 3$, $20 \leq P_0^* \leq 100$, $2 \leq Fr \leq \infty$). We identify three types of bubble collapse patterns. As an example of the first type of collapse we consider a case with $R_{\rm c}^*=0.2$, $P_0^*=50$ and $Fr=\infty$. The fact that gravity is neglected makes the behavior found in this case typical of what may be expected for relatively small bubbles. Figure \ref{Fig:4-1} shows the bubble shape at six typical moments, (a) the initial phase, (b) over-expanded phase (the gas pressure drops below the ambient pressure), (c) the maximum volume, (d) the collapse, (e) jet formation and (f) jet impact. Since  gravity is ignored in this case, there is no migration of the bubble centroid in the vertical direction. The most important feature of the bubble motion is the annular jet formation with the maximum dimensionless velocity being 2.11 (corresponding to 30 m/s for $P_\infty=2\times 10^5$Pa), which is of the same order as the jet velocity found in existing studies of bubble collapse in the vicinity of flat solid walls \cite[see e.g.][]{Philipp1998}. This similarity in order of magnitude is interesting in view of the many differences in the system geometry between these studies and ours. Note that the curvature at the equator of the bubble is the largest during most of the bubble-life, which is responsible for the formation of the annular jet. Besides, the bubble top and bottom own relatively high curvature during the collapse phase (frame d), which lead to the formation of two tiny liquid jets along the cylinder surface (frame e). In this case, the jet impacts the cylinder after the bubble reaches its minimum volume with a moderate velocity. Nevertheless, the highly-compressed bubble may cause large amplitude loads on the cylinder body, which may be a significant threat to its integrity. After the jet impact, the bubble splits into two parts, followed by the rebound of the bubble pair (frame f). The pressure amplitude decreases rapidly during the rebound phase. It is worth mentioning that even if one starts the simulation with higher curvature at the top and bottom of the bubble, the main jet we get is still an annular radial one, plus minor effects along the contact ring on the cylinder surface (not shown here). 

\begin{figure}[htbp]
	\centering\includegraphics[width=13.5cm]{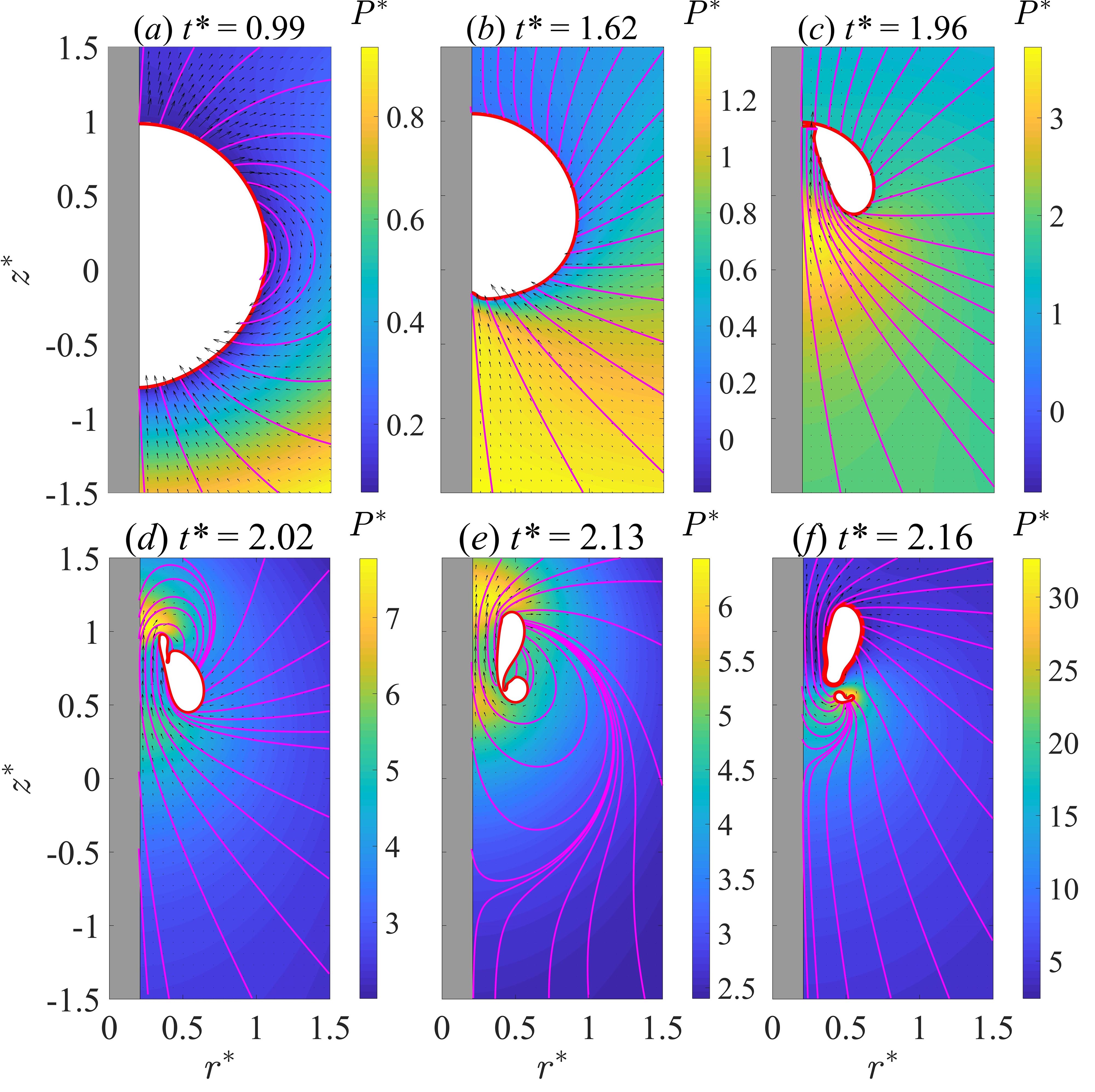}
	\caption{The collapse pattern of an attached toroidal bubble on a cylinder surface for $Fr = 2$. Other parameters are kept the same as in Figure \ref{Fig:4-1}. The bubble bottom collapses faster than other locations of the bubble surface and an upward jet forms. After the penetration of the jet, the bubble detaches from the cylinder surface and further splits into two parts, followed by the rebound of the upper daughter bubble and a further collapse of the lower one.}\label{Fig:4-2}
\end{figure}

Figure \ref{Fig:4-2} shows the collapse pattern of a bubble subjected to a strong buoyancy effect ($Fr=2$); the rest of the parameters are the same as in Figure \ref{Fig:4-1}. The expansion phase of the bubble is not shown here since it is similar to that in Figure \ref{Fig:4-1}. Gravity generates a relatively significant pressure gradient around the bubble (frame a). Therefore, the bubble bottom collapses faster (frame b) and a pronounced upward liquid jet forms afterwards (frame c). The jet speed reaches 3.5, which is higher than that in the previous case. The jet impact causes the detachment of the bubble from the cylinder surface and a high-pressure region can be observed around the jet tip (frame d). Here we adopted a vortex ring model \cite{Wang1996,WangQX2005} to simulate the interaction between the toroidal bubble and the cylinder after the detachment. The development of an annular neck on the toroidal bubble leads to the breakup of the bubble (frame e). Thereafter, the upper daughter bubble rebounds while the lower one continues to collapse for a while (frame f). A multiple-vortex-ring model  \cite{Zhang2015b,Li_oe2016} was employed here for the simulation of the breakup of a toroidal bubble.

\begin{figure}[htbp]
	\centering\includegraphics[width=14cm]{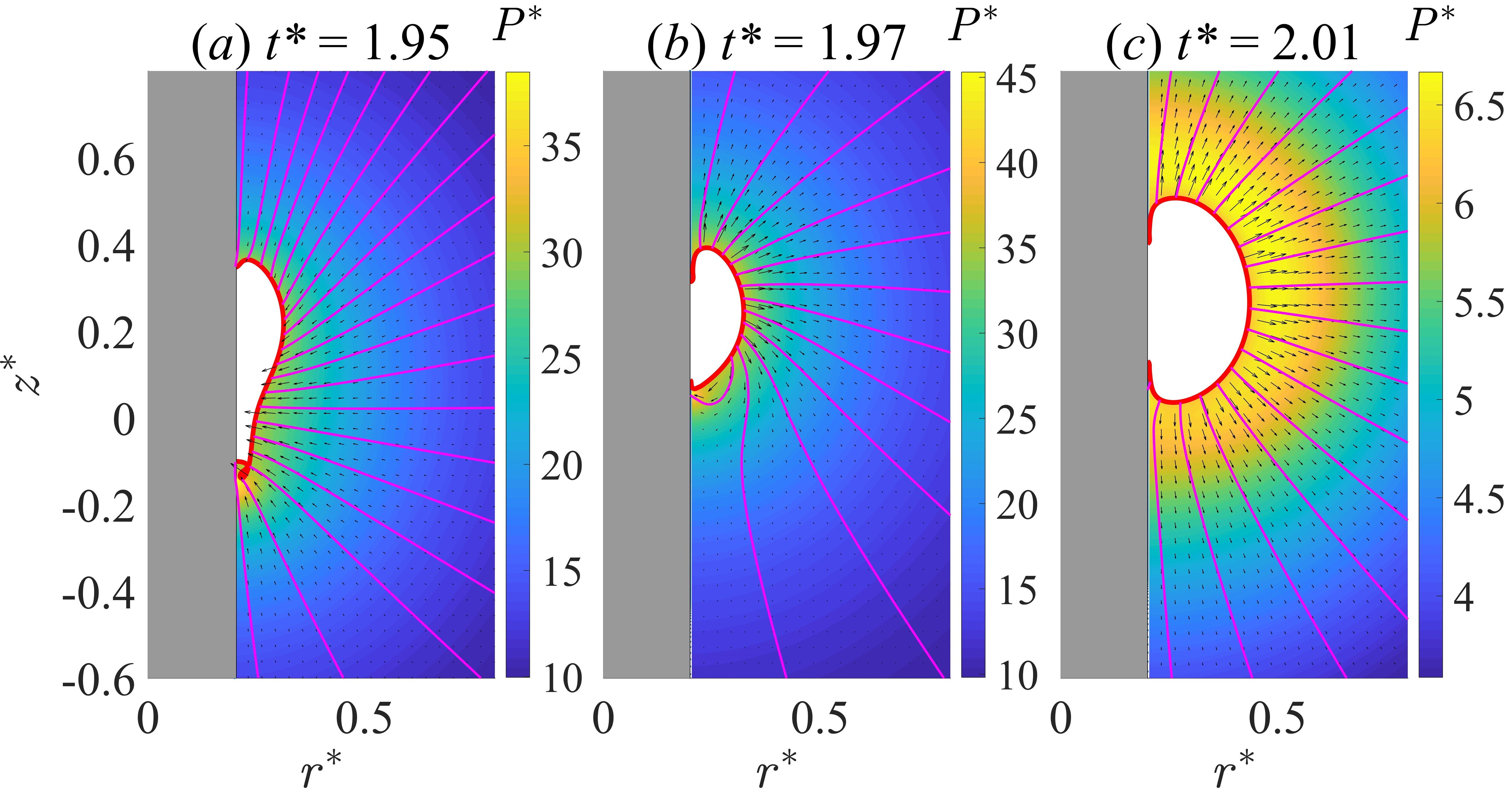}
	\caption{The collapse and rebound of an attached toroidal bubble on a cylinder surface for $Fr = 9$, i.e., the bubble is subjected to an intermediate amount of gravity. Other parameters are kept the same as in Figure \ref{Fig:4-1}. A very thin jet forms at the final collapse stage. Meanwhile, an annular jet (flow) forms from the side and collides with the upward jet, leading to the shedding of a very small bubble. Again, a thin jet forms from the bottom of the left main bubble (frame b). However, the jet cannot thread through as the bubble is rebounding (frame c). Note that the small bubble pinched in frame (a) has been removed from the simulation.}\label{Fig:4-3}
\end{figure}

These two examples demonstrate two types of jets that may appear during the bubble-cylinder interaction, namely, (i) an annular jet and (ii) an upward jet induced by gravity. One may expect that these two jets occur simultaneously when the gravity effect is weaker than that in Figure \ref{Fig:4-2}. Figure \ref{Fig:4-3} gives the collapse and rebound process of a bubble for a case of this type, for $Fr = 9$. At the final collapse stage (frame a), a very thin jet forms from the bubble bottom. Meanwhile, an annular jet/flow forms from the side of the bubble and collides with the upward jet, leading to the shedding of a very small bubble. For simplicity, we continue the simulation removing the tiny bubble. This treatment can be found in the literature \cite{Han-OE,Fong,Borkent2009} and the removal of this tiny bubble only causes less than 1\% energy loss of the system.  After the pinch-off of the tiny bubble, another thin jet forms from the bottom of the big bubble (frame b). However, the jet cannot thread through as the bubble is  rebounding (frame c). This type of bubble collapse pattern is referred to as `No jet' in this study. 

\begin{figure}[htbp]
	\centering\includegraphics[width=16cm]{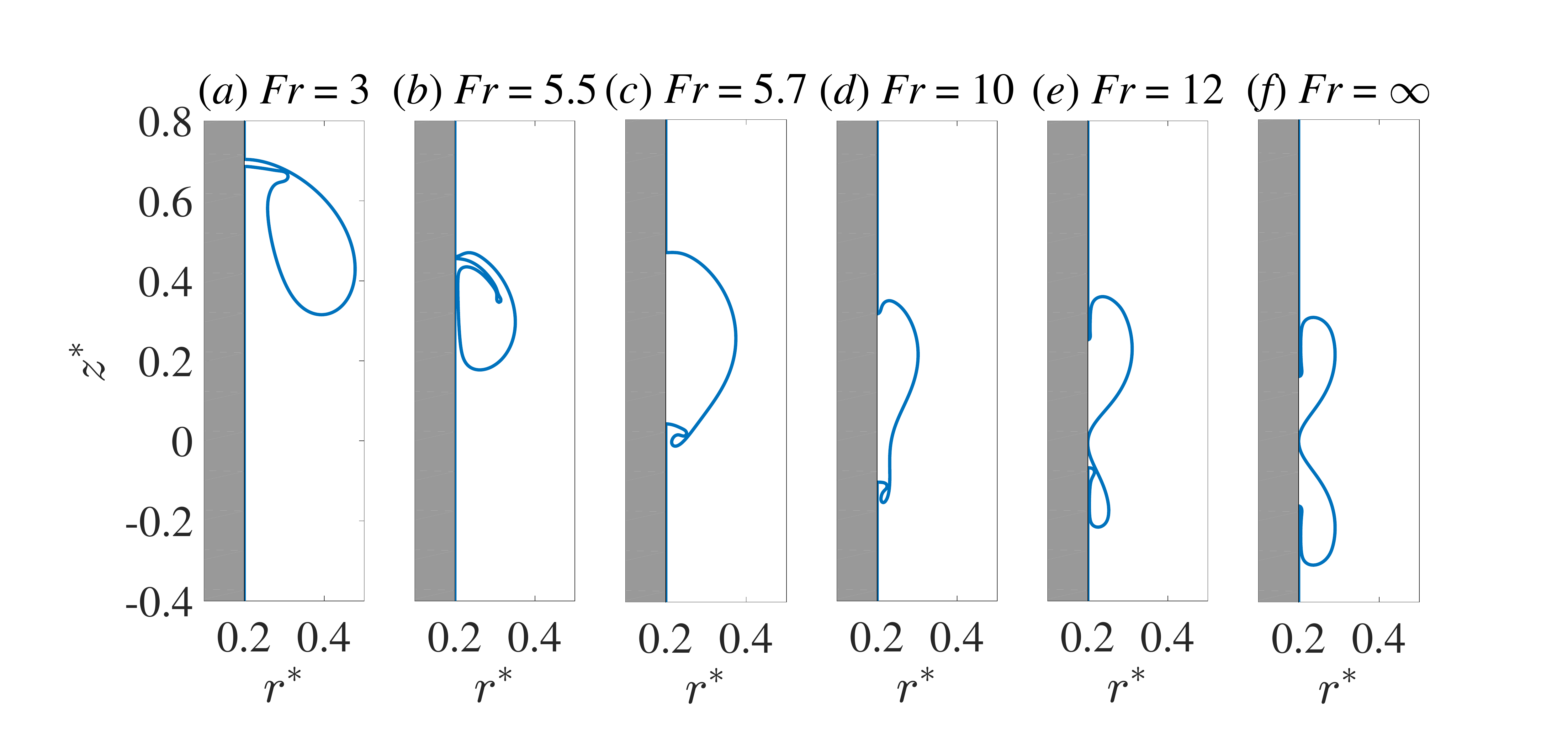}
	\caption{Bubble profiles at the jet penetration moment or pinch-off moment for different $Fr$. Other parameters are kept the same as in Figure \ref{Fig:4-1}. The dimensionless times for each frame are 1.962, 1.966, 1.934, 1.954, 1.971 and 1.973. }\label{Fig:4-4}
\end{figure}

\begin{figure}[htbp]
	\centering\includegraphics[width=15cm]{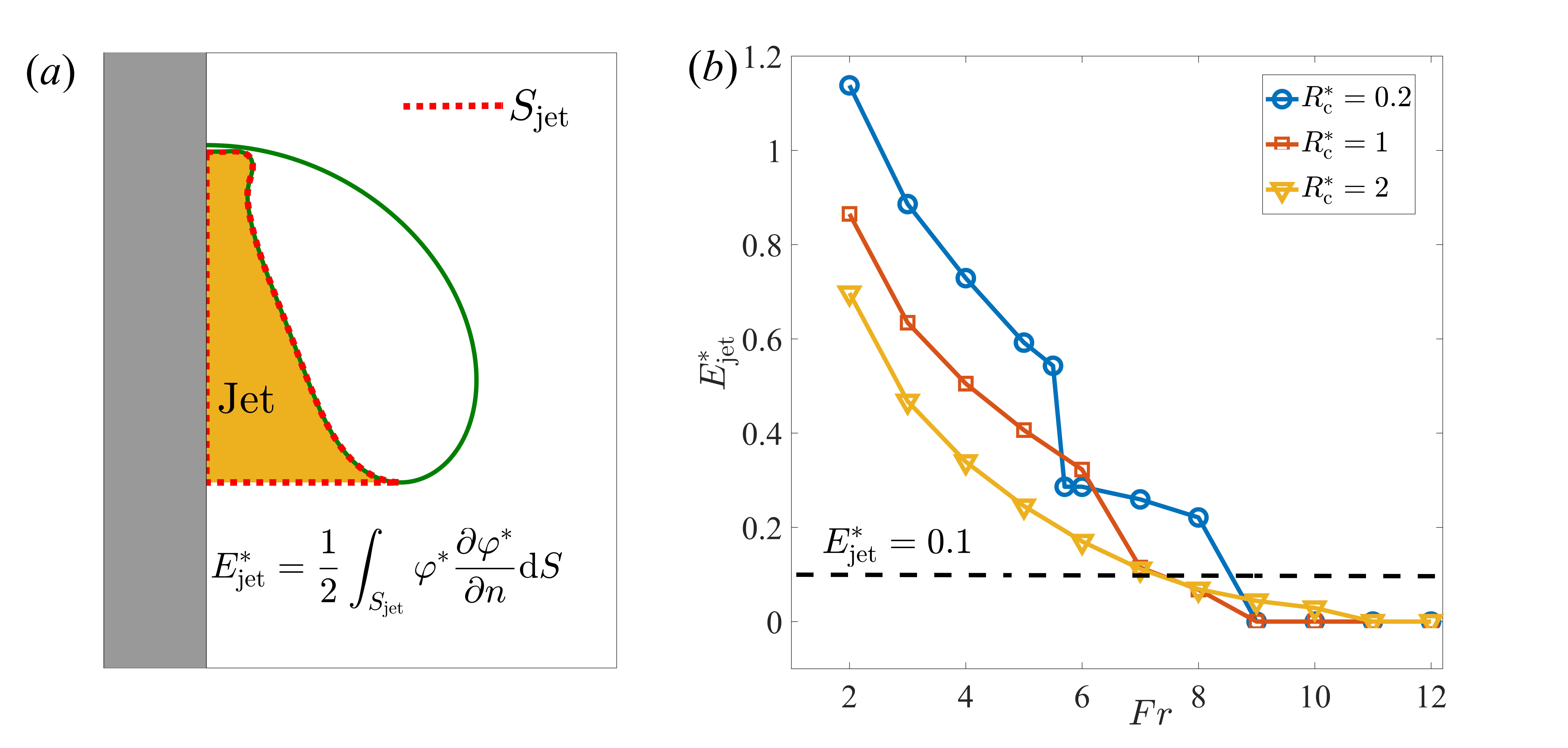}
	\centering\includegraphics[width=10cm]{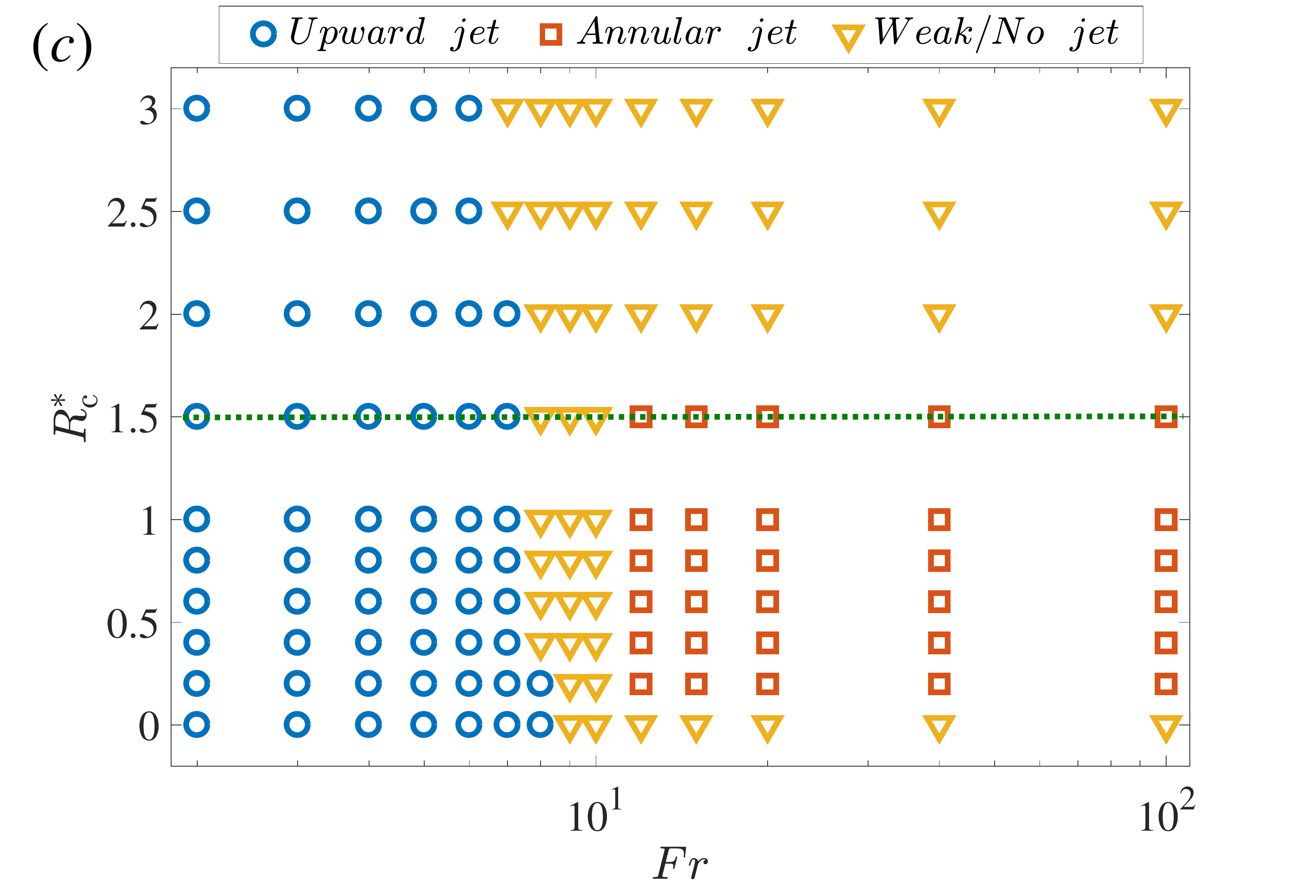}
	\caption{(a) Sketch explaining how the kinetic energy of the liquid jet is calculated. (b) Evolution of $E_{\rm jet}^*$ as a function of $Fr$ for different $R_{\rm c}^*$. The black dashed line denotes $E_{\rm jet}^*$ = 0.1, below which the jet is relatively weak. (c) Classification of different bubble collapse patterns for bubble-cylinder interactions as a function of $Fr$ and $R_c^*$, i.e., upward jet (blue circles), annular jet (red rectangles) and weak/no jet (yellow triangles). The green dashed line denotes $R_c^*$ = 1.5. Above this line, there are two collapse patterns of the bubble; Below this line, there are three patterns.}\label{Fig:three-types}
\end{figure}

Figure \ref{Fig:4-4} shows more bubble profiles at the jet impact moment (frames a-b, e-f) or at the pinch-off moment (frames c-d) for different $Fr$. The rest of the parameters are the same as in Figure \ref{Fig:4-1}. An upward jet can be expected when $Fr \lesssim 5.5$ and an annular jet can be seen when $Fr \gtrsim 12$. The shedding of the tiny bubble from the lower part of the bubble occurs between the above two regions. We further simulated the bubble motion after the tiny bubble pinch-off and found two different phenomena: (i) no jet impact (see Figure \ref{Fig:4-3}) and (ii) upward jet (not shown here). To quantify the upward jet, we introduce the kinetic energy of the liquid jet $E_{\rm jet}^*$, similar to the procedure in Pearson et al. \cite{Pearson2004-jet}. The way in which $E_{\rm jet}^*$ is calculated is shown in Figure \ref{Fig:three-types}(a). Here we only calculate the kinetic energy of a vertical jet at the impact moment. The evolution of $E_{\rm jet}^*$ over $Fr$ for different $R_{\rm c}^*$ is shown in Figure \ref{Fig:three-types}(b). For $R_{\rm c}^*$ = 2, $E_{\rm jet}^*$ decreases smoothly with $Fr$. For $R_{\rm c}^*$ = 0.2, $E_{\rm jet}^*$ decreases very fast around $Fr=5.5$ and $Fr=8$, which correspond to the transition points of different bubble collapse patterns. We found that the upward jet can be very thin in some situations with a weak influence of gravity. In this study, we use a threshold value for $E_{\rm jet}^*$ to define the `weak jet', namely, when the dimensionless kinetic energy of the jet is smaller than 0.1 (the total energy of the system is 4.94), as indicated by the black dashed line in Figure \ref{Fig:three-types} (b). In the following discussion on the phase diagram, we put the `weak jet' and `no jet' together because Li et al. \cite{Li-IJMF} demonstrated that this kind of weak jet (the ratio between $E_{\rm jet}^*$ and the total energy of the system is less than $\sim2\%$) has a negligible effect on the pressure wave induced by a free-field airgun-bubble.

After performing hundreds of simulations with different parameters, we obtain a map of the bubble collapse patterns in the $Fr-R_c^*$ space, as shown in Figure \ref{Fig:three-types}(c). Here the initial bubble pressure is fixed at $P_0^*=50$. The choice of $P_0^*$ has been tested to have little effect on the bubble collapse pattern before the jet impact. For a clear explanation of the map, we first divide the map with a horizontal line ($R_{\rm c}^* = 1.5$). When $R_{\rm c}^* \le 1.5$, there are three distinct bubble collapse patterns, as discussed above. Gravity plays a vital role when $Fr \le 7$ and an upward jet occurs. If the cylinder radius is very small ($R_{\rm c}^* \le 0.2$), an upward jet can be observed in a larger region ($Fr \le 8$). As long as the cylinder radius is not very small ($0.1 \le R_{\rm c}^* \le 1.5$), an annular jet can be observed when $Fr \ge 12$. Between the two regions, there is a relatively narrow region that presents weak or no jet. Above the horizontal line ($R_{\rm c}^* = 1.5$), we can only observe two regions, namely with an upward jet or no jet. The annular jet region disappears, which can be understood as follows: When the cylinder radius increases, the toroidal bubble becomes more slender and the curvature difference on the bubble surface becomes smaller. Therefore, the half-toroidal bubble tends to keep its shape when the influence of gravity is weak. To conclude, a very large cylinder  ($R_{\rm c}^* > 1.5$) or a very small cylinder  ($R_{\rm c}^* < 0.1$) hardly induce an annular jet of the bubble.

\subsection{The influence of the cylinder radius}
\label{S:4-3}

\begin{figure}[htbp]
	\centering\includegraphics[width=16cm]{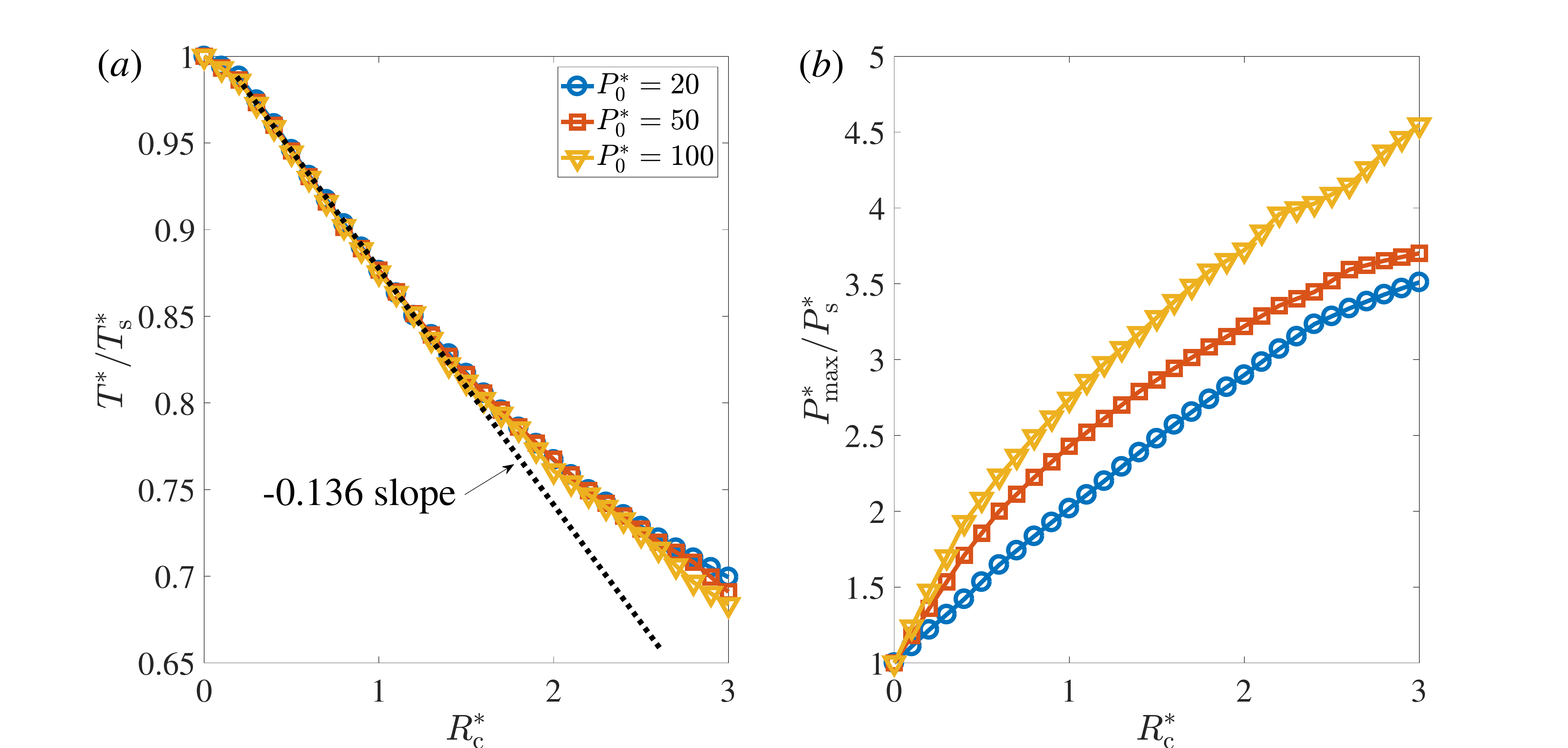}
	\caption{The influence of the cylinder radius on (a) the bubble oscillation period $T^*$ and (b) the maximum pressure induced by the bubble collapse $P\rm_{max}^*$ without the influence of gravity ($Fr=\infty$). Three series of simulations are carried out for different initial bubble pressures, namely $P_0^*=20$, 50 and 100, respectively. In this and subsequent figures, we normalize $T^*$ and $P\rm_{max}^*$ with the corresponding values for a spherical bubble (denoted by $T_{\rm s}^*$ and $P\rm_{s}^*$, respectively).}\label{Fig:Rc-effect}
\end{figure}

In this section, we aim to quantitatively investigate the influence of the cylinder radius $R_{\rm c}^*$ on the bubble oscillation period $T^*$ and the pressure peak $P_{\rm max}^*$ induced by the first bubble collapse (see Figure \ref{Fig:verification1}b for the definitions). These two quantities are crucial to the spectrum for the pressure wave. The natural frequency of the bubble can be approximated by $1/T$. Here three series of simulations are carried out for different initial bubble pressures ($P_0^*=$ 20, 50 and 100) and $R_{\rm c}^*$ ranges from 0 to 3. In  practical applications, the cylinder radius is smaller than the maximum bubble radius, i.e., $R_{\rm c}^* < 1$. Here we use an extended range of $R_{\rm c}^*$ to investigate the physics of the bubble-cylinder interactions. Figure \ref{Fig:Rc-effect} (a) shows the influence of $R_{\rm c}^*$ on the bubble oscillation period $T^*$ without gravity ($Fr=\infty$). Since different initial pressures of the bubble are used, we normalize $T^*$ with the corresponding value of a spherical bubble, which are given by $T_s^*=$ 2.079, 1.987 and 1.944 for $P_0^*$ being 20, 50 and 100, respectively. Clearly, the bubble oscillation period is not very sensitive to the initial bubble pressure and only slightly decreases with increasing $P_0^*$. All curves almost lie on top of each other, especially in the range of $0 \le R_{\rm c}^* \le 2$. The bubble oscillation period decreases with $R_{\rm c}^*$, which means the bubble oscillates faster with a larger $R_{\rm c}^*$ and the resonant frequency of the bubble increases. One important reason for this phenomenon is that the semi-toroidal bubble becomes more slender as $R_{\rm c}^*$ increases. Keep in mind that the initial energy of the bubble is the same in all simulations for a specified $P_0^*$. Therefore, the bubble period tends to zero when $R_{\rm c}^*$ approaches infinity. Compared to a spherical bubble, the reduction of $T^*$ is about 12\% when $R_{\rm c}^*=1$ and less than 5\% when $R_{\rm c}^* \le 0.4$. The plot also reveals a linear dependency of the bubble oscillation period on $R_{\rm c}^*$ in the range of $0.2 \le R_{\rm c}^* \le 1.5$ with the slope being -0.136. 

\begin{figure}[htbp]
	\centering\includegraphics[width=16cm]{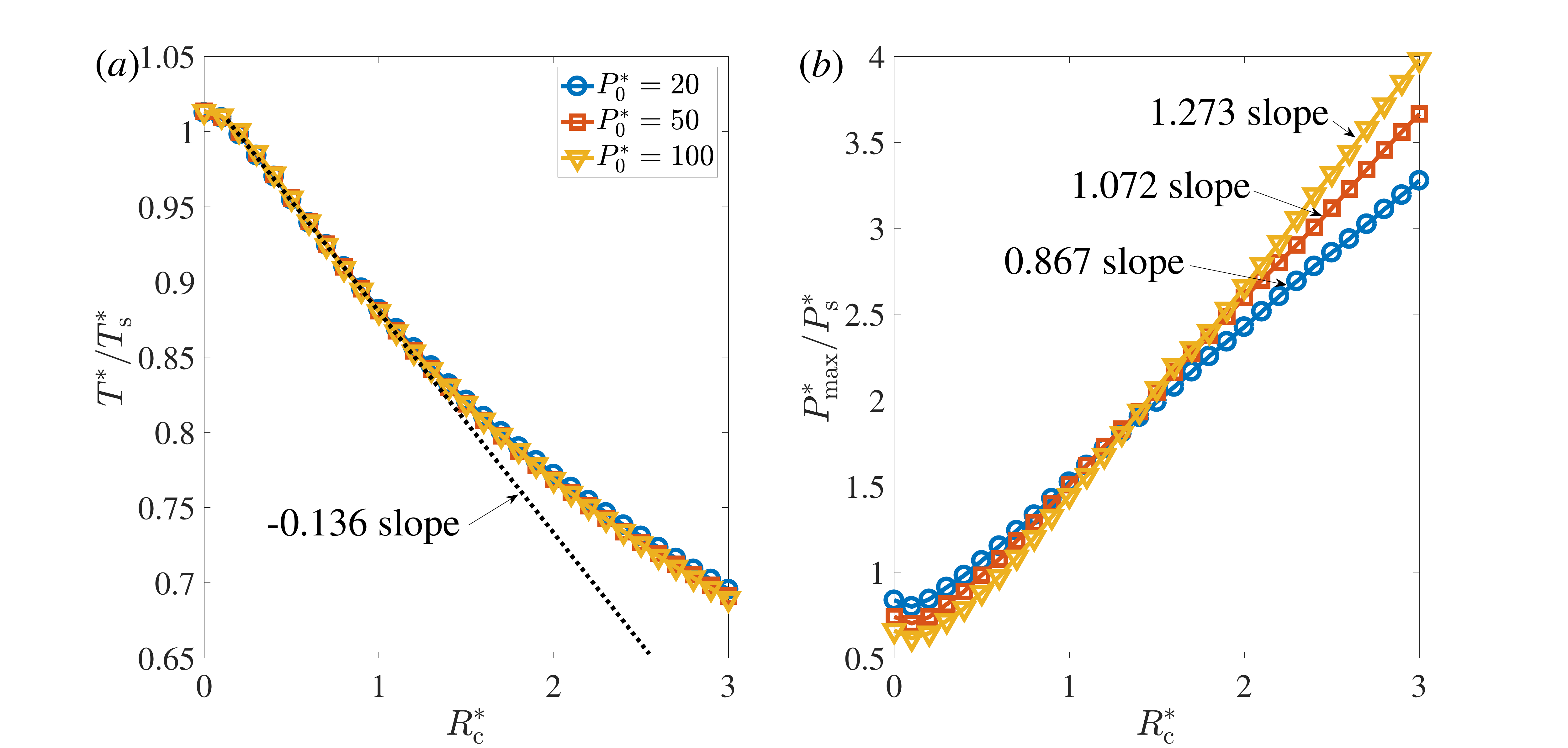}
	\caption{The influence of the cylinder radius on $T^*$ and $P\rm_{max}^*$ with a stronger gravitational component ($Fr=3$). The rest of the parameters are kept the same as in Figure \ref{Fig:Rc-effect}.}\label{Fig:Rc-effectFr3}
\end{figure}

Figure \ref{Fig:Rc-effect} (b) shows the variation of the maximum pressure induced by the bubble collapse versus $R_{\rm c}^*$. Here we also normalize $P_{\rm max}^*$ with the corresponding value of a spherical bubble ($P_{\rm s}^* = $ 5.579, 10.330 and 16.345 for $P_0^*$ = 20, 50 and 100, respectively, where as expected, $P_{\rm s}^*$ increases with increasing $P_0^*$). $P_{\rm max}^*/P_{\rm s}^*$ increases with $R_{\rm c}^*$, which is connected to the fact that the maximum bubble radius of the cross section is smaller with increasing $R_{\rm c}^*$ due to geometric reasons. Therefore, the bubble dynamics is faster and since the far-field pressure is proportional to the second derivative of the bubble volume, $P_{\rm max}^*/P_{\rm s}^*$ increases. The increase rate of $P_{\rm max}^*/P_{\rm s}^*$ generally slows down with $R_{\rm c}^*$. In addition, for a fixed finite $R_c^*$, $P_{\rm max}^*/P_{\rm s}^*$ increases slightly with $P_0^*$.

Starting from the above results, we further carried out simulations with a strong gravity effect ($Fr=3$) whereas other parameters are the same as in Figure \ref{Fig:Rc-effect}. As discussed above, an upward jet forms when the bubble is subjected to strong gravity. The evolution of  $T^*/T_{\rm s}^*$ and $P_{\rm max}^*/P_{\rm s}^*$ as functions of $R_{\rm c}^*$ are given in Figure \ref{Fig:Rc-effectFr3}. As for $T^*/T_{\rm s}^*$ (see panel a), the characteristics of the curves are quite similar to those without gravity (see Figure \ref{Fig:Rc-effect}a). We also notice that there is a region ($0.2 \le R_{\rm c}^* \le 1.3$) where the $T^*/T_{\rm s}^*$ decreases linearly with $R_{\rm c}^*$ and the slope is the same as that in Figure \ref{Fig:Rc-effect}(a).  Compared with the zero-gravity situation ($Fr=\infty$),  $T^*/T_{\rm s}^*$ slightly increases when the bubble is subjected to a strong influence of gravity, which may be attributed to the fact that the ambient pressure surrounding the bubble decreases during the upward migration of the bubble.

Figure \ref{Fig:Rc-effectFr3}(b) shows the dependence of $P_{\rm max}^*/P_{\rm s}^*$ on $R_{\rm c}^*$. For a thin cylinder ($R_{\rm c}^* \le 0.2$), $R_{\rm c}^*$ has little effect on $P_{\rm max}^*/P_{\rm s}^*$. In this case, due to gravity, a vigorous liquid jet forms with a width larger than the cylinder radius. The surface of the bubble detaches from the cylinder surface (at the jet impact moment) and positions itself relatively far from the cylinder (e.g., Figure \ref{Fig:4-2}). Hence, the collapse pattern of bubbles in this region have much similarity, in terms of the pressure field and the bubble oscillation period (see Figure \ref{Fig:Rc-effectFr3}a). This finding is useful for practical applications. Specifically, for large-scale airgun bubbles, gravity dominates the jetting behavior and the airgun-body plays a minor role when $R_{\rm c}^* \le 0.2$. Generally, under this condition, $T^*/T_{\rm s}^*$ and $P_{\rm max}^*/P_{\rm s}^*$ differ by less than 2\% from the case  $R_{\rm c}^*=0$. Starting from $R_{\rm c}^*=0.2$, $P_{\rm max}^*/P_{\rm s}^*$ increases linearly with $R_{\rm c}^*$ and the slope increases with $P_0^*$, which shows quite different features compared with the gravity-free situation (see Figure \ref{Fig:Rc-effect}b). Finally, note that $P_{\rm max}^*/P_{\rm s}^*$ starts below 1 since for a free bubble the influence of gravity decreases the magnitude of the pressure peak, but then quickly increases to values above 1 for $R_{\rm c}^* \gtrsim 0.5-0.7$.

\subsection{The influence of the Froude number}
\label{S:4-4}

In this section, we quantitatively investigate the influence of the Froude number $Fr$ (defined by Eq. \ref{Equation:Fr}) with the cylinder size fixed at $R_{\rm c}^*=0.2$ again using different initial bubble pressures $P_0^*=$ 20, 50 and 100; $Fr$ ranges from 2 to infinity. For $Fr<2$, the influence of gravity is very strong and the bubble moves upward rapidly so that the surrounding hydrostatic pressure falls and the collapse is weak. 

\begin{figure}[htbp]
	\centering\includegraphics[width=16cm]{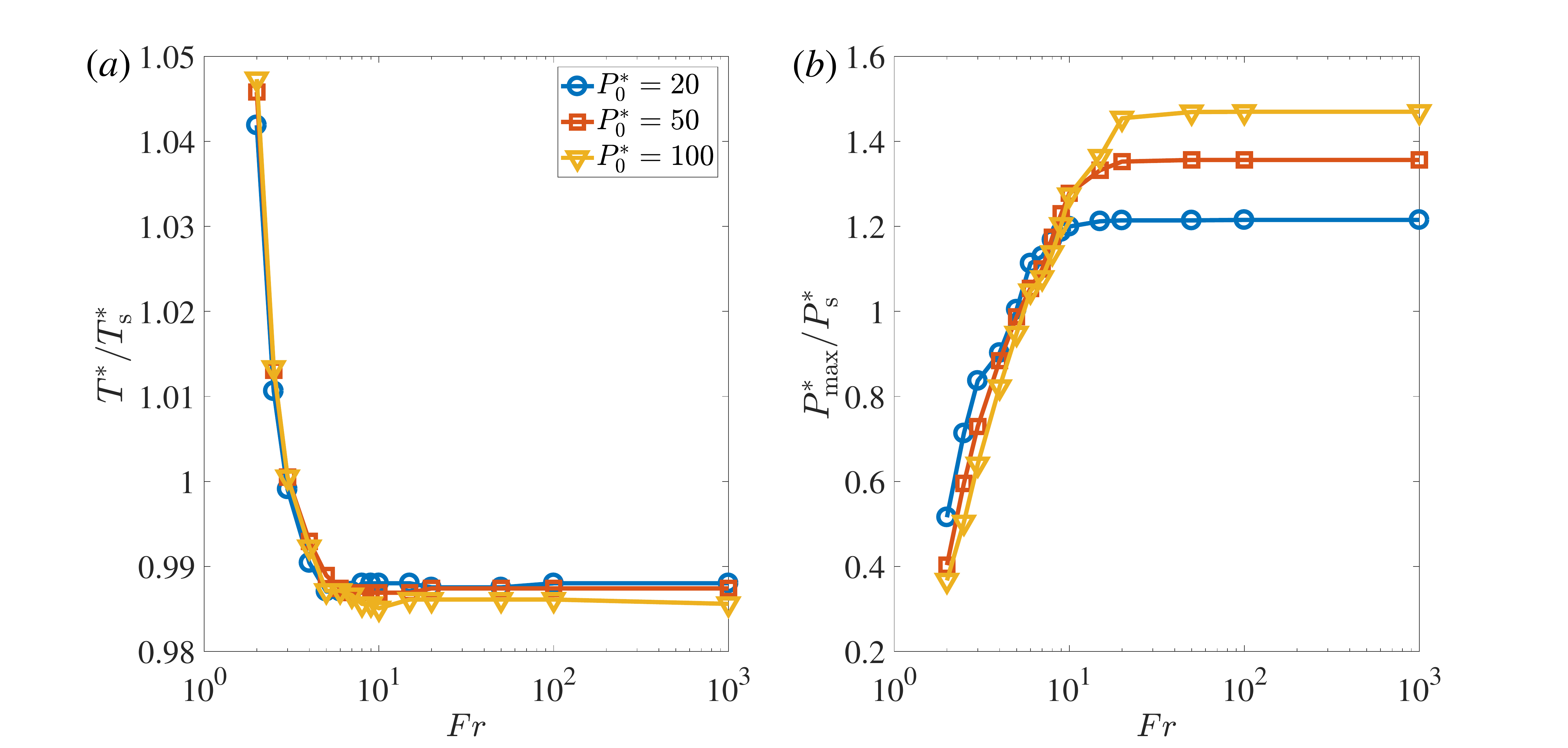}
	\caption{The influence of the Froude number $Fr$ on $T^*$ and $P\rm_{max}^*$ for a fixed cylinder radius $R_c^* =0.2$ .}\label{Fig:effect-Fr}
\end{figure}

Figure \ref{Fig:effect-Fr}(a) shows the evolution of the bubble oscillation period as a function of $Fr$. $T^*/T_{\rm s}^*$ decreases by about 6\% when $Fr$ increases from 2 to 10 and remains constant when $Fr>10$. As already discussed above, $T^*/T_{\rm s}^*$ is not strongly affected by the initial bubble pressure. Turning to $P_{\rm max}^*/P_{\rm s}^*$ (see Figure \ref{Fig:effect-Fr}b), we observe an increase as the influence of gravity becomes smaller and the bubble rises less and less when $Fr$ becomes sufficiently large. This finding is in consistent with our previous study on a single airgun-bubble without a cylinder \cite{Li-IJMF}. The detailed explanation of the mechanism behind this behavior can be found therein.

\section{The impact of the cylinder on realistic airgun-bubbles}
\label{S:5}

All the simulations in Section \ref{S:4} started with a high-pressure gas bubble, and the gas release phase of a real airgun-bubble was not considered. In a previously published paper \cite{Li-IJMF} which neglected the presence of the airgun body, we proposed a model to consider the process of air release, liquid compressibility \cite{wang_blake_2010,wang_blake_2011} and heat transfer \cite{Chelminski,Graaf2014}.  In this section, we aim to use this model to investigate the influence of the cylinder on a realistic airgun-bubble. The object of study is a Sercel type 520 airgun, for which the experimental data can be found in de Graaf et al \cite{Graaf2014}. The radius of this airgun is around 0.15 m and its length 0.64 m. Other parameters of the airgun were set to the real values, including the chamber volume (8521 cm$^3$), the air pressure (17.2 MPa) and the port area (128 cm$^2$). Additionally, the opening time of the airgun valve was set to 4 ms according to Li et al \cite{Li-IJMF}. Since the length of this airgun is small,the expanding bubble encloses the airgun body quickly during the expansion \cite{Graaf2014,Graaf2014b}. However, to simplify the simulation, we treated the airgun-body as a long cylinder. In order to gain some insight into the effect of the finite length of the real airgun body. We conducted simulations with $R_c$ = 0, 0.15 and 0.30 m.

\begin{figure}[htbp]
	\centering\includegraphics[width=7.7cm]{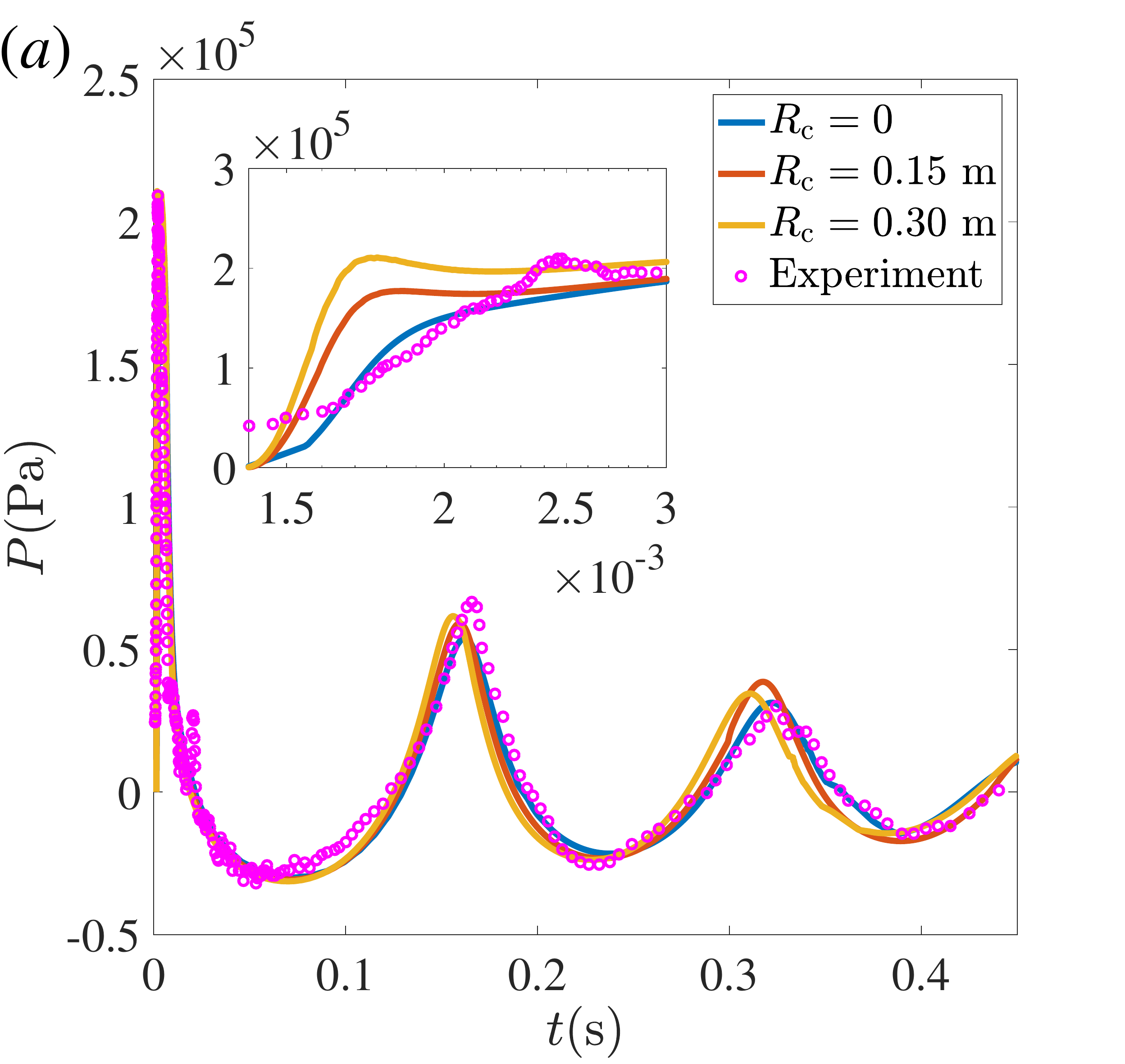}
	\centering\includegraphics[width=7.7cm]{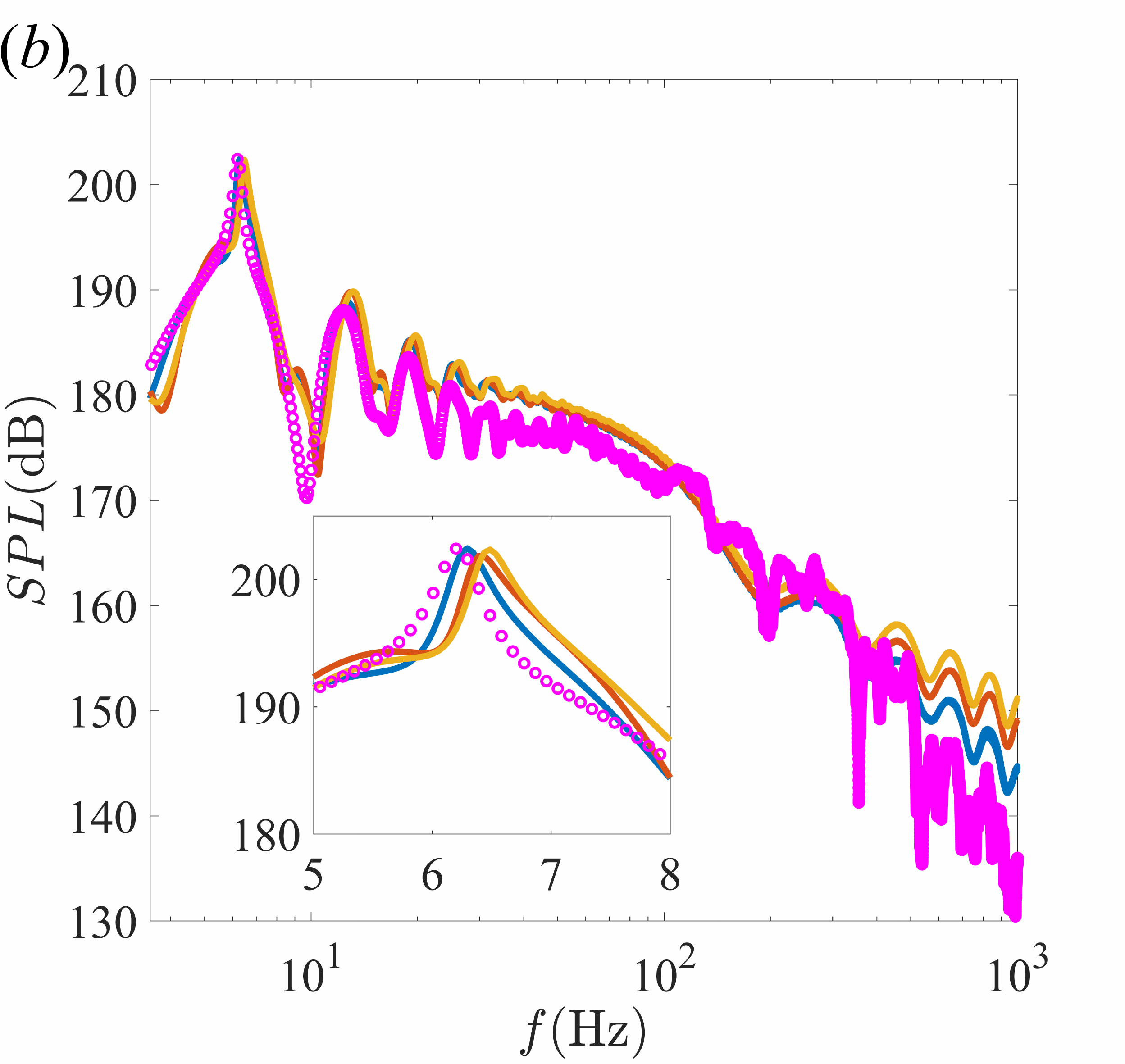}
	\caption{The impact of a cylinder on (a) the pressure waves and (b) the associated spectrums for a realistic airgun-bubble. The pressure sensor was installed at the same depth as the airgun and 2.22 m away. The experimental date can be found in de Graaf et al \cite{Graaf2014}.}\label{Fig:real-airgun}
\end{figure}

Figure \ref*{Fig:real-airgun} (a) shows a comparison of the experiment and numerical pressure waves. In the experiment, the pressure sensor was installed at the same depth as the airgun and 2.22 m away. Keep in mind that the results discussed in this section are dimensional and the pressure signal was calculated from Equation (\ref{eq:pressure}), where $r$ was set according to the experiment. The numerical results shown here have been superposed with the reflected signals from the free surface with the reflection coefficient being -1 \cite{Graaf2014,Cox}. The bubble oscillation period slightly decreases with increasing $R_{\rm c}$ and the pressure peak slightly increases, which are consistent with the findings of Section \ref{S:4}. Since the expanding bubble encloses the airgun-body quickly during the expansion \cite{Graaf2014,Graaf2014b}, the bubble motion is little affected by the presence of the airgun-body. Therefore, the numerical results match the experiment better for $R_{\rm c}$ = 0.

Figure \ref*{Fig:real-airgun} (b) shows a comparison of the pressure waves spectrum. The highest peak corresponds to the natural frequency of the bubble (see the inset). A larger airgun-body $R_{\rm c}$ leads to an increase of the natural frequency, which is not desired in practical applications. Additionally, the high-frequency content becomes stronger with increasing $R_{\rm c}$. The numerical results meet the experimental data well in the low-frequency range ($f<20$ Hz), especially the highest peak of the curve. The high-frequency content seems to be overestimated, which may be attributed to the limited resolution of the experimental data we measured from the printed figure \cite{Graaf2014}.

The high-frequency waves (10-150 kHz) not only are useless in deepsea geophysical exploration but also disturb, injure or kill marine life \cite{Landr2011,Khodabandeloo2018}. To reduce the environmental impact, the high-frequency content should be controlled below a safety level. Many studies have shown that the rise time of the first pressure peak is an indicator of high-frequency content generated \cite{Watson2019,Chelminski,Li-IJMF}. As can be seen in the inset of Figure \ref*{Fig:real-airgun} (a), a decreasing $R_{\rm c}$ also decreases the slope of the first peak and the high-frequency content.
Recently, Watson et al \cite{Watson2019} found that the air release rate and consequently the first pressure wave depend on the airgun length. Specifically, due to the influence of the rarefaction wave inside the chamber, a longer airgun causes a slower decay of the first pressure wave and hence reduced high-frequency content. Therefore, we recommend that the environmentally friendly airgun of the future should have a relatively smaller radius and a longer body in a reasonable range.

\section{Summary and conclusions}
\label{S:con}
This paper systematically investigated the strong interaction between a vertical cylinder with an attached toroidal bubble via hundreds of boundary integral (BI) simulations. Such a configuration is associated with an application to geophysical exploration, namely seismic survey. Firstly, the BI code was validated by comparison to an analytical solution, convergence tests and comparison to an experiment. Subsequently, we discussed the influence of the main governing parameters (the cylinder radius $R_{\rm c}^*$ normalised by the maximum equivalent bubble radius, initial bubble pressure $P_0^*$ normalised by the ambient pressure and Froude number $Fr$ defined in Equation \ref{Equation:Fr}) on the bubble collapse pattern, oscillation period and pressure wave. And finally, we studied the impact of the presence of a cylinder on a realistic airgun-bubble with particular focus on the spectrum of the pressure waves. The main findings are:

\begin{enumerate}[\indent(1)]
	
	\item When $R_c^* \le 1.5$, three types of bubble collapse patterns are identified for different combination of $Fr$ and $R_{\rm c}^*$, namely (i) upward jetting due to gravity, (ii) annular jet toward the cylinder body and (iii) no jet. When $R_{\rm c}^* > 1.5$, only two types of bubble collapse patterns remain and the annular jet region disappears.	See Figure \ref{Fig:three-types}(c) for a map of the bubble collapse patterns in the $Fr-R_{\rm c}^*$ space.
	
	\item The bubble oscillation period $T$ decreases slightly with $R_{\rm c}^*$. Compared with a spherical bubble, the reduction of $T$ is slightly above 12\% when $R_c^*=1$ and less than 5\% when $R_c^*\le 0.4$. When $Fr=3$, $T/T_{\rm s}$ (with $T_s$ the Rayleigh period of a spherical bubble) decreases linearly with $R_{\rm c}^*$ in the range of $0.2 \le R_{\rm c}^* \le 1.3$ with a slope of -0.136. As $Fr$ increases, the linear relation between $T/T_{\rm s}$ and $R_{\rm c}^*$  is well verified in a larger range of $R_{\rm c}^*$ and the slope almost keeps the same.

	\item For large-scale airgun bubbles subjected to a strong influence of gravity, the effect of the airgun-body is negligible if $R_{\rm c}^* \le$ 0.2. The pressure wave can be approximated by neglecting the airgun  with an error within  2\%. Starting from $R_{\rm c}^*=0.2$, the ratio of the maximum pressure wave $P_{\rm max}$ to that of a spherical bubble $P_{\rm s}$ increases linearly with $R_{\rm c}^*$ and the slope increases with $P_0^*$. 
	
	\item For a relatively thin cylinder $R_{\rm c}^*$ = 0.2, $T/T_{\rm s}$ decreases by $\sim$ 6\% when $Fr$ increases from 2 to 10 and remains constant when $Fr\ge 10$. $P_{\rm max}/P_{\rm s}$ increases rapidly when $Fr$ increases from 2 to 10 and then keeps a constant.
	
	\item A large cylinder body has considerable effects on the spectrum of the pressure wave induced by a realistic airgun bubble. As $R_{\rm c}$ increases, the high-frequency content increases and the low-frequency content decreases. Further, considering the effect of the chamber length on the air release rate \cite{Watson2019}, we recommend that the environmentally friendly airgun of the future should have a relatively smaller radius and larger length in a reasonable range.
	
\end{enumerate}

\section{Acknowledgements}
\label{S:ack}

The authors gratefully acknowledge Detlef Lohse for insightful discussions. This work was funded by SHELL.





\bibliographystyle{model1-num-names}
\bibliography{sample}







\end{document}